# On the use of Standardized Precipitation Index (SPI) for drought intensity assessment

M. Naresh Kumar[a*], C.S. Murthy[b], M.V.R. Sesha Sai[b] and P.S. Roy[b]

[a] Software Development & Database Systems Group

[b] Remote Sensing & GIS Applications Area

National Remote Sensing Centre, Hyderabad 500 625, India

## Abstract

Monthly rainfall data from June to October for 39 years was used to generate Standardized Precipitation Index (SPI) values based on Gamma distribution for a low rainfall and a high rainfall district of Andhra Pradesh state, India. Comparison of SPI, with actual rainfall and rainfall deviation from the mean indicated that SPI values under-estimate the intensity of dryness/wetness when the rainfall is very low/very high respectively. As a result, the SPI in the worst drought years of 2002 and 2006 in the low rainfall district has indicated only moderate dryness instead of extreme dryness. The range of SPI values of the high rainfall district indicated better stretching, compared to that of the low rainfall district. Further, the SPI values of longer time scale (2-, 3- and 4- months) showed an extended range compared to 1-month, but the sensitivity in drought years has not improved significantly.

To ascertain whether non normality of SPI is a possible reason, normality tests were conducted. The Shapiro-Wilk statistic, p-values and absolute value of the median confirmed normal distribution of SPI in both the districts whereas cumulative probability distribution of SPI indicated deviation from normal probability in the lower and upper ranges.

Therefore, it is suggested that SPI as a stand alone indicator needs to be interpreted with caution to assess the intensity of drought. Further investigations should include; sensitivity of SPI to the estimated shape and scale at lower and upper bounds of gamma and impact of other distributions such as Pearson III on SPI computation, to complement the above results.

---

[*] Corresponding author – mail address nareshkumar_m@nrsa.gov.in

# 1. Introduction

Meteorological drought is the earliest explicit event in the process of occurrence and progression of drought. Rainfall is the primary driver of meteorological drought. There are numerous indicators based on rainfall that are being used for drought monitoring (Smakhtin and Hughes, 2007). Rainfall deviation from normal -a long term mean, is the most commonly used indicator for drought monitoring. In India, on the basis of rainfall deviations, four categories namely ±20% deviation as normal, -20 to -60% deviations as deficit, -60% and below as scanty, above 20% as excess are being used for evaluating the rainfall patterns across the country during the monsoon season (www.imd.gov.in). The declaration of meteorological drought is done if the total season's rainfall is less than 75% of long term mean, with -50 to -74% deviations representing moderate drought and less than -50% deviations representing severe drought (www.imd.gov.in). In South Africa, less than 70% of normal precipitation is considered as drought and such a situation for two consecutive years indicate severe drought (Bruwer, 1990). In Poland also, rainfall deviation from multi year mean (equivalent to long term mean) forms the criterion for drought monitoring in Poland (www.imgw.pl).

Although rainfall deviation from mean continues to be a widely adopted indicator for drought intensity assessment because of its simplicity, the application of this indicator is strongly limited by its inherent nature of mean dependence. Rainfall deviations cannot be applied uniformly on different areas having varying mean rainfall. A high rainfall area and low rainfall area can have the same rainfall deviation for two different amounts of actual rainfall. Therefore, rainfall deviations across space and time need to be interpreted with due care.

Standardized Precipitation Index (SPI) expresses the actual rainfall as a standardized departure with respect to rainfall probability distribution function and hence the index has gained importance in recent years as a potential drought indicator permitting comparisons across space and time. The computation of SPI requires long term data on precipitation to determine the probability distribution function which is then transformed to a normal distribution with mean zero and standard deviation of one. Thus, the values of SPI are expressed in standard deviations, positive SPI indicating greater than median precipitation and negative values indicating less than median precipitation (Edwards and McKee, 1997). Since SPI values fit a typical normal distribution, these values lie in one standard deviation approximately 68% of time, with in 2 sigma 95% of time and with in 3 sigma 98 % of time. In recent years SPI is being used increasingly for assessment of drought intensity in many countries (Vijendra *et al.,* 2005; Wu *et al.,* 2006; Vicente-Serrano *et al.*, 2004). The homogeneous climatic zones were derived using SPI in Mexico (Giddings *et al.,* 2005). Time series analysis of SPI indicated decrease in SPI values during 1970-1999 reflecting the increase in dry conditions in southern Amazon region (Li *et al.*, 2007). The drought interpretation at different time scales using SPI is proved to be superior to Palmer Drought Index (Guttman, 1998). Yearly values of Palmer Drought Severity Index and SPI were used to rank the years according to drought severity by Goodrich and Ellis, 2006. Smakhtin and Hughes, 2007, developed software to compute and apply different rainfall based indicators for quantitative assessment of meteorological drought. McKee *et al.* 1993, suggested the SPI ranges for different severity levels of drought (Table I).

The present study analyses the response of seasonal SPI values to drought situation vis-à-vis comparison of SPI with actual rainfall and rainfall deviation from normal in a low rainfall and a high rainfall district. The main objective is to investigate whether SPI can perform as a better indicator for drought intensity assessment than conventional and widely adopted rainfall deviations.

SPI computation with monthly rainfall data using two parameter Gamma distribution, analysis of SPI in relation to rainfall deviation from mean, interpretation of SPI values to detect dry ness and wetness in drought and normal years, study of the impact of record length on SPI, evaluation of normality tests for SPI and issues for further investigation constitute the outline of the current research paper.

## 2. Study area and methodology

Two districts of Andhra Pradesh state, India, namely, Ananthpur representing low rainfall and Khammam representing high rainfall were selected. The total geographic area of Ananthpur district is 19135 sq km and that of Khammam district is 15809 sq. km. Monthly actual rainfalls and corresponding normals from June to October for 39 years (1969 to 2007), collected from the Directorate of Economics and Statistics, Government of Andhra Pradesh, India, was used as input data in the analysis. The rainfall pattern of the two districts shown in (Table II) indicates that Ananthpur district has season's total normal i.e., long term average rainfall of 449mm, whereas Khammam district has 997mm incident rainfall. Ananthpur district has been declared by the state administration as chronic drought prone area because of its low rainfall with high inter-annual variability. In India, there are 185 districts in 13 states, occupying 120 m ha of geographic area identified as drought prone areas (Murthy *et al.,* 2008). Khammam district is not a drought prone because of its stable and higher rainfall pattern. Thus, the two districts with contrasting rainfall patterns were selected for analyzing the behavior of SPI.

Computation of SPI with the time series data, at monthly scale, was done based on two parameter Gamma distribution function. The computation of SPI involves transformation of precipitation data into lognormal values followed by computation of U statistics, shape and scale parameters of the gamma distribution. The resulting parameters are then used to find the incomplete gamma cumulative probability of an observed precipitation event. The incomplete gamma cumulative probability is then converted to gamma probabilities after including the occurrences of zero precipitation events. The gamma probabilities are then transformed in to standardized normal distribution using equi-probability transformation techniques (Abramowitz and Stegun 1965). Although the transformation can be achieved through analytical methods we employed a statistical method following Edwards and Mc Kee, 1997. The detailed computation procedure was furnished in the appendix.

Rainfall deviation from normal was calculated using the formula ((Actual rainfall-Normal rainfall)/Normal rainfall)*100, which expresses the actual rainfall as percent deviation from normal. The normal rainfall is the long term average of the actual rainfall.

## 3. Results and discussion

The analysis of the current study is focused on understanding the sensitivity of SPI to actual rainfall/rainfall deviation and the behavior of SPI in drought and normal years. The SPI based drought classes proposed by Mc Kee *et al*. 1993, have been adopted in this study (Table I), because of its wider applicability to different regions of climatology such as Mexico (Giddings, 2005), Greece (Loukas *et al.,* 2004), Iran (Morid *et al.,* 2006), European Alps (Bartolini *et al.,* 2008), Portugal (Paulo *et al.,* 2005), Europe (Llyod *et al.,* 2002), Poland (Łabędzki *et al*., 2005), mountainous Mediterranean basin (Vicente-Serrano *et al*., 2004), Slovenia (Ceglar *et al.,* 2008), Colorado, North Dakota, Iowa, Kansas, Nebraska, South Dakota, and Wyoming ( Wu *et al*., 2006), Eastern China (Bordi *et al.,* 2004), Northeast of Thailand (Wattanakij *et al*., 2006), South Africa (Rouault, 2003). As suggested by McKee et al. 1993, SPI represents wetter and drier climates in a similar way.

Scatter plots of SPI versus deviation from normal rainfall for July and August months were drawn for positive rainfall deviation (i.e., actual rainfall is greater than normal) and for negative rainfall deviation (i.e., actual rainfall is less than normal) as shown in (Figures 1 and 2). July and August months are very critical from an agriculture point of view. July rainfall is very critical for crop sowings and August rainfall is vital for the growth of different crops. Rainfall pattern in these two months plays a greater role in the occurrence of agricultural drought. It is evident from Figures 1 and 2 , particularly in the low rainfall district – Ananthpur, that very high negative deviations (-60 to -80%) representing very low rainfall events are associated with SPI values of -1.00 to -1.50 in most of the cases despite the fact that such a severe dryness should correspond to the SPI of -2.00 and below. Similarly, rainfall deviations of -40 to -60 % of normal, which is quite significant reduction from normal rainfall correspond to SPI values of -0.5 to -1.0 indicating mild dryness or less significant dryness. In the high rainfall district – Khammam, the SPI values are on lower side compared to low rainfall district, for the higher negative rainfall deviations.  Thus, there is a relation between SPI and rainfall deviations, but the magnitude of SPI values does not indicate the severity of drought situation.

Positive rainfall deviations, indicating that actual rainfall is more than normal are associated with positive SPI values indicating wetness, in both the months.  But, the extent of positive deviation did not commensurate with the extent of positive values of SPI indicating the degree of wetness. The rainfall deviation of 50 to 100% implies that actual rainfall is 150 to 200% of normal has resulted in the SPI values of 0.5 to 1.00 signifying normal or slightly wet situation. The deviation from 100 to 200% of normal rainfall has resulted in SPI of around 1.5 indicating moderate wetness in Ananthpur district. Again, in the high rainfall district – Khammam, the SPI values tend to be on higher side >2.0, for the events of excess rainfall.

Actual values of very low and very high rainfall events and associated SPI values are shown in (Table III) to bring more clarity on the inter relations between SPI and rainfall.  Even the very small amounts of rainfall that is certainly not enough to maintain enough soil moisture for agriculture have resulted in the SPI values of around -1.5 which otherwise should represent extreme dryness with the values around -2.0 and below. Similarly, excess rainfall events have the SPI around 1.5. This trend of very low rainfall events not resulting in a very low SPI and very high rainfall events not resulting in a very high SPI was evident in all the five months.

Therefore, from the foregoing analysis, it is clearly evident that the SPI values are over estimated for low rainfall levels and underestimated for high rainfall levels, in the study area districts, particularly in the low rainfall district. In the high rainfall district the values of SPI are more stretched between +2.0 to -2.0, with better agreement with actual rainfall situation compared to that of lower rainfall district.

### 4. SPI of drought and normal years

In the study area district – Ananthpur, 2002 and 2006 are the worst drought years and 2000 is a normal year as declared by the State administration. Peanut (Arachis Hypogea) is the principal crop with more than 80 percent of cultivated area. The intensity of drought situation is understood from the statistics published by the Government which reads that the yield of groundnut crop was 67 kg/ha in 2006, 355 kg/ha in 2002 and 1118 kg/ha in 2000. Comparison of SPI and rainfall deviations, pertaining to drought years and normal year, was studied to understand the sensitivity of SPI and its agreement with rainfall deviations (Figures 3 and 4). Both SPI and rainfall deviations exhibit the same trend with normal year at higher level and two drought years falling much lower to normal. The rainfall deviations are very significant, ranging from -40 to -80% reflecting the deficiency in most of the months in 2002 and 2006 and signifying severe drought situation. Positive rainfall deviations indicating excess rainfall in most of the months signify the normal season in the year 2000.

The values of SPI in the drought year 2002, ranged between 0 to -0.1 in most of the months. In the drought year 2006, SPI was lowest at -1.5 in August, around -1.0 in July and October and around -0.05 in September. By applying SPI classes corresponding to drought intensity proposed by McKee et al. 1993, the worst drought years of 2002 and 2006 in the study area district represent only mild to moderate drought situation. Thus, the drought intensity was underestimated by SPI based classes, mainly due to the over estimation of SPI values at very low rainfall events as discussed in previous section. Even in the good year like 2000, which had recorded the highest groundnut crop yield, the SPI values are around 1.0 indicating normal situation, as a result of underestimation of SPI at high rainfall events.

### 5. Longer time scale

The longer time scale, 2-, 3- and 4- months' rainfall data is used for computing the SPI to understand its behavior with respect to 1-month SPI. The comparison of SPI and rainfall deviations is carried out for Ananthpur district (Figure 5). The values of SPI are -2 and below for rainfall deviations less than -50% and the SPI tend to be greater than 2 for the high rainfall events. Thus, the range of SPI values is higher with stretching beyond -2 and +2, for longer time scale SPI compared to 1-month SPI and thus longer the time scale of SPI, higher is the range. Longer time scale SPI values during drought and normal years were shown in (Table IV), which indicate that even in drought years of 2002 and 2006 the SPI values are around -1.5 indicating only moderate dryness.

## 6. Record length

The impact of variable record length on the SPI was studied by considering different time periods from 21 years (1969-1989), 22 years (1969-1990), 23 years (969-1991) and so on upto 39 years (1969-2007) of data for the two study area districts and for July and August months separately. SPI calculation for each incremental year from the initial 21 years period 1969-1989 till 2007, resulted in 19 SPI values. Maximum and minimum SPI were identified from these 19 values of each month and were plotted separately for each district and month as shown in Figure 6. The negligible difference between maximum and minimum SPI as the record length increases from 21 years (corresponding period is 1969-1989) to 39 years (corresponding period is 1969-2007), indicates that the SPI is stable and not influenced by the length of record. As a result, SPI based interpretation on different events of dryness/wetness remains consistent. The results are in agreement with the findings of the study by Wu et al. 2005. This property of SPI suggests the robustness of the indicator, particularly when the analysis of very long term rainfall data is involved.

## 7. Agreement of results with earlier studies

The results of the present study are in agreement with the findings of the earlier studies to some extent. Wu et al. 2006, revealed that the application of SPI of short time scales in arid and the areas with distinct dry season fails to detect the occurrence of drought situation. This behavior of SPI is attributed to its non normal distribution caused by higher frequency of no rain cases. Histograms of drought frequency classes derived by Morid *et al*. 2006, showed that percent normal rainfall has higher frequency in extreme drought and severe drought, where as SPI have higher frequency in normal class. The result indicated that for the cases of low percent normal rainfall which represents lower and lower rainfall, the corresponding SPI values tend to be higher indicating normal situation.

Interpretation of 1-month SPI can lead to misleading assessment, as there are many examples with small rainfall deviations leading to large positive or negative SPI values. Actual precipitation of 15.2 mm against the normal of 2.5 mm leads to SPI of +3.11. Similarly 371.9 mm of precipitation which is above the normal by 211.6 mm, gave rise to SPI value of 1.97. In another station, 24.9 mm of precipitation against 10.4 mm of normal which is 239% of normal, has resulted in the SPI value of 1.43. February 1996 SPI of -1.76 over Southeastern Plains Climate Division in New Mexico represents zero rainfall situations (http://www.drought.unl.edu/monitor).

## 8. Tests of normality

Thus, non normal distribution caused by the occurrence of zero rainfall events was found to be responsible for the distorted SPI values in low and uncertain rainfall areas by Wu *et al*., (2006). However, in the present study area districts, there is no zero rainfall in the data set. Three tests of normality suggested by Wu *et al*. 2006, i.e., Shapiro-Wilk statistic, p-values and absolute value of the median were carried out to verify the normality of SPI. The calculated values of these three parameters are shown in (Table V). A non normal distribution should have w value less than 0.96, p value less than 0.10 and median > 0.05. By, applying the criteria, it was found that the SPI values for all the months conform to the normal distribution in both the study area districts.

Normal probability of SPI and its comparison with standard normal probability for July and August months, for two districts separately, was shown in (Figures 7 and 8). It could be observed that the SPI probability is deviating from normal line in the lower ranges and upper ranges of SPI in both the districts. Non normality observed in these two specific ranges of SPI is incidentally associated with the under estimation or over estimation of SPI as revealed in previous sections. The normality of SPI is not fulfilled in all ranges of SPI although majority of SPI values run close to the normality line. As a result, it may be required to undertake normality tests in different ranges of SPI.

## 9. Summary and Conclusion

The actual rainfall expressed as a percent deviation from normal (long term average) is the most commonly used drought indicator, although it has limited use for spatial comparison due to its dependence on mean. Standardized Precipitation Index (SPI) expresses the actual rainfall as a standardized departure with respect to rainfall probability distribution function and hence the index has gained importance in recent years as a potential drought indicator permitting comparisons across different rainfall zones.

In this study, the SPI values of different years are analyzed with actual rainfall and rainfall deviation from normal in a low rainfall and drought prone district. The objective is to evaluate whether SPI can be used as a better indicator than conventionally adopted rainfall deviation based approach for drought intensity assessment.

Scatter plots of rainfall deviations vs. SPI indicated less sensitivity of SPI to low rainfall events. A very low or very high rainfall has not corresponded to a very low (-2.0 or less) or very high (+2.0 or more) SPI values. Thus, SPI values under estimated the dryness or wetness when the rainfall is very low or very high respectively.

As a result, the worst drought years of 2002 and 2006 in the study area district represent only moderate dryness based on SPI classes proposed by McKee *et al.* 1993. SPI values of the high rainfall district indicated enhanced range of values, -2.0 or less for very low rainfall and +2.0 or more for high rainfall, compared to the low rainfall district. To ascertain whether non normality of SPI is a possible reason, normality test was conducted for SPI values based on Shapiro-Wilk statistic, p-values and absolute value of the median as suggested by Wu *et al*. 2006, and the results confirmed normal distribution of SPI in both the districts. However, visual inspection of normal probability plot of SPI indicated deviation from normal line in the lower and higher ranges of SPI values. Thus, non normality was observed in the selective ranges of SPI.

Thus, the results of the present study suggest that SPI as a stand alone indicator needs to be interpreted with caution for drought intensity assessment particularly in low rainfall districts which are more vulnerable to droughts.

Although the statistical nature of SPI permits comparisons across space and time better than rainfall deviations, the drought intensity at a given location is found to be more sensitive to rainfall deviations than SPI.

Since rainfall and its variations are very critical in low rainfall districts, SPI values should assume wider range to represent the degree of wetness or dryness to result in better assessment of drought situation. In this context, the use of other distributions such as Pearson-III distribution as suggested by Guttman (1999), for SPI computation needs to be investigated for improving the sensitivity of SPI. Further, the impact of shape and scale at lower and upper bound of gamma estimate on SPI is also an important issue that needs to be investigated.

## 10. Acknowledgements

We express our sincere thanks to Dr. V. Jayaraman, Director, National Remote Sensing Centre for his constant encouragement and guidance. Thanks are also due to Dr. R.S. Dwivedi, Group Director, Land Resources Group, for his suggestions. The cooperation offered by the Officials of Directorate of Economics and Statistics, Government of Andhra Pradesh, India for providing the data needed for the study is duly acknowledged.

**Appendix for Computation of SPI**

Procedure and Formula for Computation of SPI
1. The transformation of the precipitation value in to standardized precipitation index has the purpose of
    a. Transforming the mean of the precipitation value adjusted to 0
    b. Standard deviation of the precipitation is adjusted to 1.0
    c. Skewness of the existing data has to be readjusted to zero

    When these goals have been achieved the standardized precipitation index can be interpreted as mean 0 and standard deviation of 1.0
2. Mean of the precipitation can be computed as

$$Mean = \overline{X} = \frac{\sum X}{N} \quad \text{(A1)}$$

Where N is the number of precipitation observations
In EXCEL the mean is computed as Mean=AVERAGE (first:last)

3. The standard deviation for the precipitation is computed as

$$s = \sqrt{\frac{\sum (X - \overline{X})^2}{N}} \quad \text{(A2)}$$

In EXCEL the standard deviation is computed as s=stdevp(first:last)

4. The skewness of the given precipitation is computed as

$$skew = \frac{N}{(N-1)(N-2)} \sum \left(\frac{X - \overline{X}}{s}\right)^3 \quad \text{(A3)}$$

5. The precipitation is converted to lognormal values and the statistics U, shape and scale parameters of Gamma distribution are computed.

$$\log mean = \overline{X}_{\ln} = \ln(\overline{X}) \quad \text{(A4)}$$

$$U = \overline{X}_{\ln} - \frac{\sum \ln(X)}{N} \quad \text{(A5)}$$

$$shape\,parameter = \beta = \frac{1 + \sqrt{1 + \frac{4U}{3}}}{4U} \quad \text{(A6)}$$

$$scale\, parameter = \alpha = \frac{\overline{X}}{\beta} \quad (A7)$$

The Equations A1 to A8 is computed using built functions provided by EXCEL software.

The resulting parameters are then used to find the cumulative probability of an observed precipitation event. The cumulative probability is given by:

$$G(x) = \frac{\int_0^x x^{a-1} e^{\frac{-x}{\beta}} dx}{\beta^\alpha \Gamma(\alpha)} \quad (A8)$$

Since the gamma function is undefined for x=0 and a precipitation distribution may contain zeros, the cumulative probability becomes:

$$H(x) = q + (1-q)G(x) \quad (A9)$$

Where q is the probability of zero

The cumulative probability H(x) is then transformed to the standard normal random variable Z with mean zero and variance of one, which is the value of the SPI following Edwards and Mc Kee (1997); we employ the approximate conversion provided by Abromowitz and Stegun (1965) as an alternative

$$Z = SPI = -\left(t - \frac{c_0 + c_1 t + c_2 t^2}{1 + d_1 t + d_2 t^2 + d_3 t^3}\right) \quad 0 < H(x) \leq 0.5$$

$$Z = SPI = +\left(t - \frac{c_0 + c_1 t + c_2 t^2}{1 + d_1 t + d_2 t^2 + d_3 t^3}\right) \quad 0.5 < H(x) \leq 1 \quad (A10)$$

Where

$$t = \sqrt{\ln\left(\frac{1}{H(x)^2}\right)} \quad 0 < H(x) \leq 0.5$$

$$t = \sqrt{\ln\left(\frac{1}{(1.0 - H(x))^2}\right)} \quad 0.5 < H(x) \leq 1.0 \quad (A11)$$

$$c_0 = 2.515517$$
$$c_1 = 0.802583$$
$$c_2 = 0.010328$$
$$d_1 = 1.432788 \quad (A12)$$
$$d_2 = 0.189269$$
$$d_3 = 0.001308$$

The values of $c_0$, $c_1$, $c_2$, $d_1$, $d_2$, $d_3$ given in Equation (A12) are constants being widely employed for SPI computation (Abramowitz and Stegun 1965).

The SPI computation is shown for Ananthpur district for July Rainfall of 39 years starting from 1969 to 2007. The mean of precipitation is adjusted from 61.981 to SPI mean of -0.0111. The standard deviation of 52.2187 is adjusted to a standardization of 0.99760 and skewness in the data is reduced from 2.2196 to 0.766445.

| Statistics | Rainfall | ln | gamma | t transform | SPI |
|---|---|---|---|---|---|
| Mean(A1) | 61.981 | 4.12684 (A4) | (A9) | (A11) | -0.0111 |
| Standard Deviation(A2) | 52.2187 | | | | 0.99760 |
| Skewness(A3) | 2.2196 | | | | 0.766445 |
| U(A5) | | 0.2846 | | | |
| Shape(A6) | | 1.90981 | | | |
| Scale(A7) | | 32.4544 | | | |

| Year | Rainfall | lograinfall | gamma | T Transform |
|---|---|---|---|---|
| 1969 | 28 | 3.3322 | 0.2375 | 1.6956 |
| 1970 | 30 | 3.4012 | 0.2611 | 1.6388 |
| 1971 | 29 | 3.3673 | 0.2493 | 1.6668 |
| 1972 | 14 | 2.6391 | 0.0825 | 2.2338 |
| 1973 | 27 | 3.2958 | 0.2257 | 1.7253 |
| 1974 | 50 | 3.9120 | 0.4843 | 1.2042 |
| 1975 | 124 | 4.8203 | 0.9052 | 2.1706 |
| 1976 | 39 | 3.6636 | 0.3657 | 1.4183 |
| 1977 | 69 | 4.2341 | 0.6526 | 1.4541 |
| 1978 | 77 | 4.3438 | 0.7089 | 1.5711 |
| 1979 | 47 | 3.8501 | 0.4533 | 1.2579 |
| 1980 | 34 | 3.5264 | 0.3081 | 1.5345 |
| 1981 | 78 | 4.3567 | 0.7154 | 1.5854 |
| 1982 | 58 | 4.0604 | 0.5612 | 1.2835 |
| 1983 | 29 | 3.3673 | 0.2493 | 1.6668 |
| 1984 | 137 | 4.9200 | 0.9317 | 2.3167 |
| 1985 | 75 | 4.3175 | 0.6956 | 1.5424 |
| 1986 | 27 | 3.2958 | 0.2257 | 1.7253 |

| 1987 | 10 | 2.3026 | 0.0469 | 2.4736 |
| 1988 | 158 | 5.0626 | 0.9602 | 2.5393 |
| 1989 | 280 | 5.6348 | 0.9985 | 3.6144 |
| 1990 | 42 | 3.7377 | 0.3994 | 1.3549 |
| 1991 | 19 | 2.9444 | 0.1342 | 2.0041 |
| 1992 | 55 | 4.0073 | 0.5334 | 1.2346 |
| 1993 | 41 | 3.7136 | 0.3883 | 1.3756 |
| 1994 | 45 | 3.8067 | 0.4321 | 1.2955 |
| 1995 | 112 | 4.7185 | 0.8724 | 2.0292 |
| 1996 | 56 | 4.0254 | 0.5428 | 1.2510 |
| 1997 | 10 | 2.3026 | 0.0469 | 2.4736 |
| 1998 | 112 | 4.7185 | 0.8724 | 2.0292 |
| 1999 | 38 | 3.6376 | 0.3544 | 1.4405 |
| 2000 | 55 | 4.0073 | 0.5334 | 1.2346 |
| 2001 | 21 | 3.0445 | 0.1564 | 1.9262 |
| 2002 | 21 | 3.0534 | 0.1585 | 1.9193 |
| 2003 | 41 | 3.7062 | 0.3849 | 1.3819 |
| 2004 | 108 | 4.6821 | 0.8593 | 1.9806 |
| 2005 | 145 | 4.9767 | 0.9443 | 2.4033 |
| 2006 | 21 | 3.0445 | 0.1564 | 1.9262 |
| 2007 | 55 | 4.0146 | 0.5371 | 1.2412 |

**Table I Drought categories from SPI (Source: Mc Kee et al., 1993)**

| SPI | Drought category |
|---|---|
| 0 to -0.99 | Mild drought |
| -1.00 to -1.49 | Moderate drought |
| -1.5 to -1.99 | Severe drought |
| -2.00 or less | Extreme drought |

**Table II Rainfall pattern in the study area districts**

| Month | District wise normal rainfall (mm) | |
|---|---|---|
| | Ananthpur | Khammam |
| June | 64 | 132 |
| July | 67 | 314 |
| August | 89 | 280 |
| September | 118 | 165 |
| October | 111 | 106 |
| Total | 449 | 997 |

**Table III Very low rainfall events not associated with a very low SPI and very high rainfall not associated with a very high SPI**

| Month | Year | Actual rainfall (mm) | Rainfall deviation from normal % | SPI |
|---|---|---|---|---|
| June | 1988 | 13 | -72 | -1.694 |
| | 1984 | 15 | -68 | -1.532 |
| | 2004 | 18 | -72 | -1.316 |
| | 2001 | 19 | -70 | -1.250 |
| | 1987 | 84 | 79 | 1.057 |
| | 1991 | 131 | 179 | 2.007 |
| | 2007 | 141 | 130 | 2.175 |
| | 1996 | 145 | 209 | 2.247 |
| July | 1997 | 10 | -82 | -1.676 |
| | 1972 | 14 | -74 | -1.389 |
| | 1991 | 19 | -66 | -1.107 |
| | 1984 | 137 | 158 | 1.489 |
| | 2005 | 145 | 116 | 1.592 |
| | 1988 | 158 | 182 | 1.753 |
| | 1989 | 280 | 400 | 2.977 |
| August | 1972 | 6 | -92 | -2.263 |
| | 1984 | 13 | -84 | -1.671 |
| | 2004 | 15 | -83 | -1.562 |
| | 1969 | 156 | 98 | 1.308 |
| | 1998 | 166 | 131 | 1.418 |
| | 2000 | 171 | 92 | 1.471 |
| September | 1969 | 27 | -80 | -2.028 |
| | 1994 | 30 | -75 | -1.908 |

| | 2003 | 46 | -61 | -1.384 |
|---|---|---|---|---|
| | 1974 | 231 | 75 | 1.299 |
| | 2001 | 244 | 107 | 1.418 |
| | 1988 | 265 | 117 | 1.602 |
| | 1981 | 283 | 114 | 1.753 |
| October | 1976 | 39 | -58 | -1.495 |
| | 1997 | 43 | -55 | -1.366 |
| | 1991 | 197 | 105 | 1.234 |
| | 1989 | 208 | 124 | 1.353 |
| | 2001 | 226 | 104 | 1.541 |
| | 1975 | 248 | 167 | 1.757 |

**Table IV SPI of longer time scales in drought and normal years**

| Year | Situation on ground | June+ July | July+ August | August+ September | June+ July+ August | July+ August+ September | June to September |
|---|---|---|---|---|---|---|---|
| 2000 | Normal | 0.352 | 1.126 | 0.538 | 1.293 | 0.403 | 0.522 |
| 2002 | Drought | -1.173 | -1.2 | -1.424 | -1.407 | -1.635 | -1.717 |
| 2006 | Drought | -0.145 | -1.85 | -1.358 | -1.063 | -1.582 | -1.135 |

**Table V Measured values of parameters for testing normality of SPI from June to October in the study area districts**

| District | Parameters Measured | June | July | August | September | October |
|---|---|---|---|---|---|---|
| Anantpur | w value | 0.954 | 0.956 | 0.973 | 0.973 | 0.961 |
| | p value | 0.170 | 0.170 | 0.165 | 0.165 | 0.168 |
| | median | 0.023 | 0.171 | 0.117 | 0.109 | 0.085 |
| Khammam | w value | 0.948 | 0.978 | 0.966 | 0.969 | 0.981 |
| | p value | 0.172 | 0.164 | 0.167 | 0.166 | 0.163 |
| | median | 0.041 | 0.003 | 0.182 | 0.188 | 0.057 |

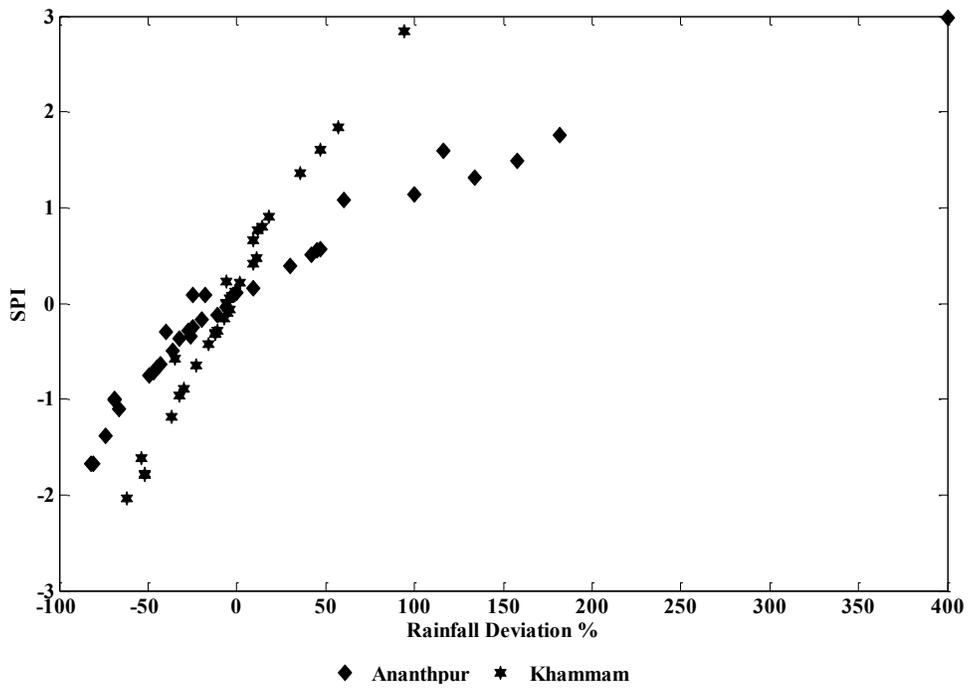

**Figure 1 Scatter plots of SPI versus deviation from normal rainfall for July**

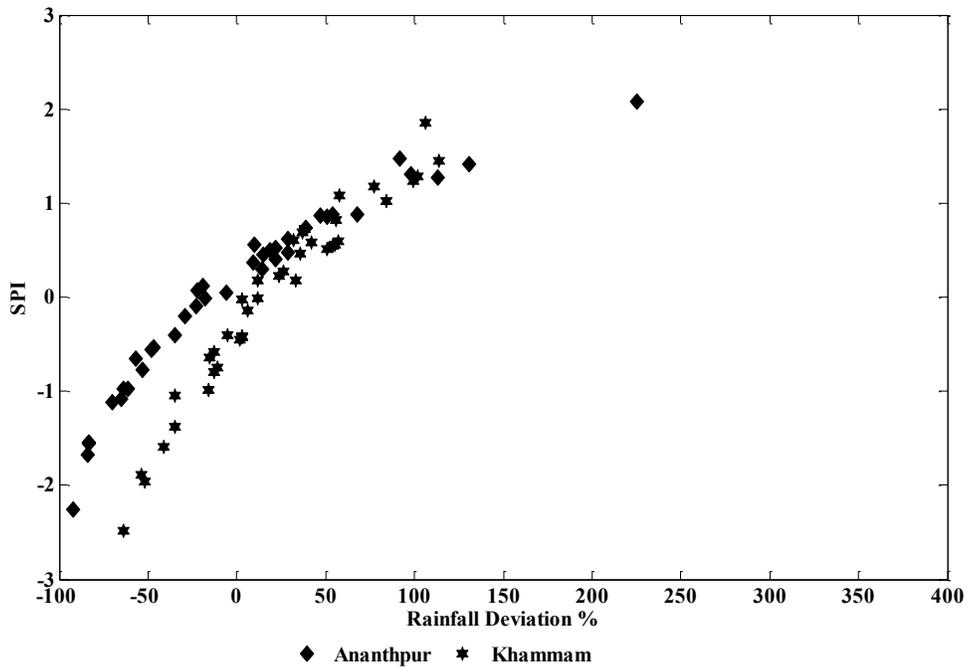

**Figure 2 Scatter plots of SPI versus deviation from normal rainfall for August**

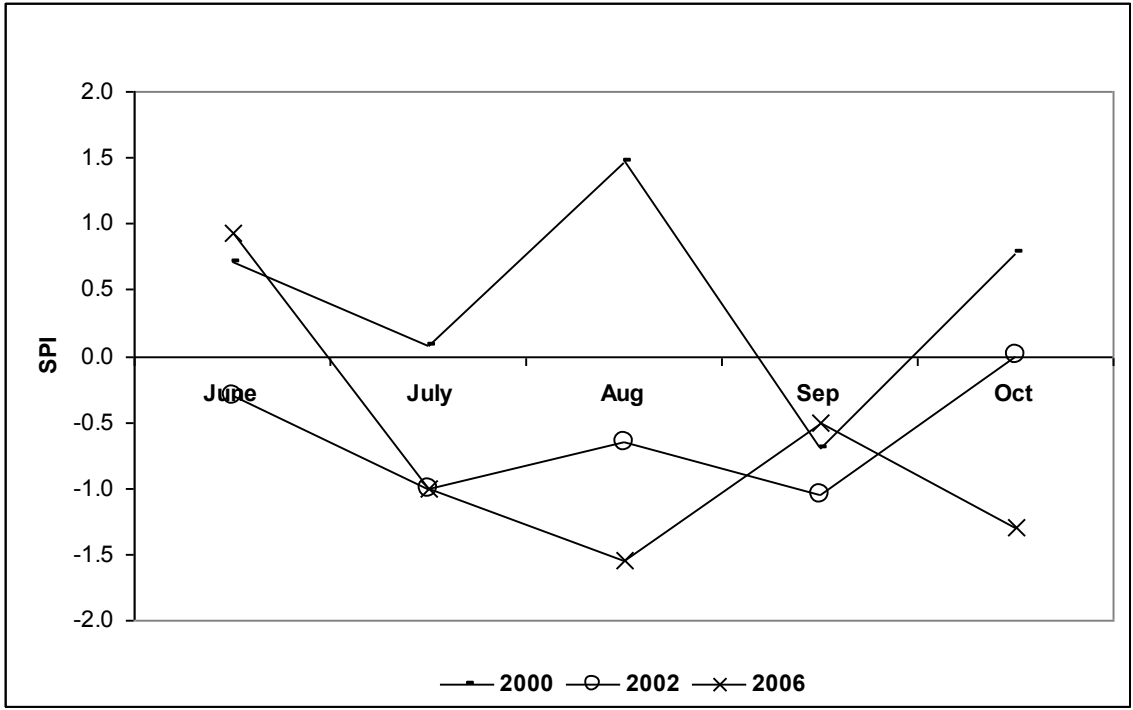

**Figure 3 SPI from June to October for drought years (2002 and 2006) versus normal year (2000)**

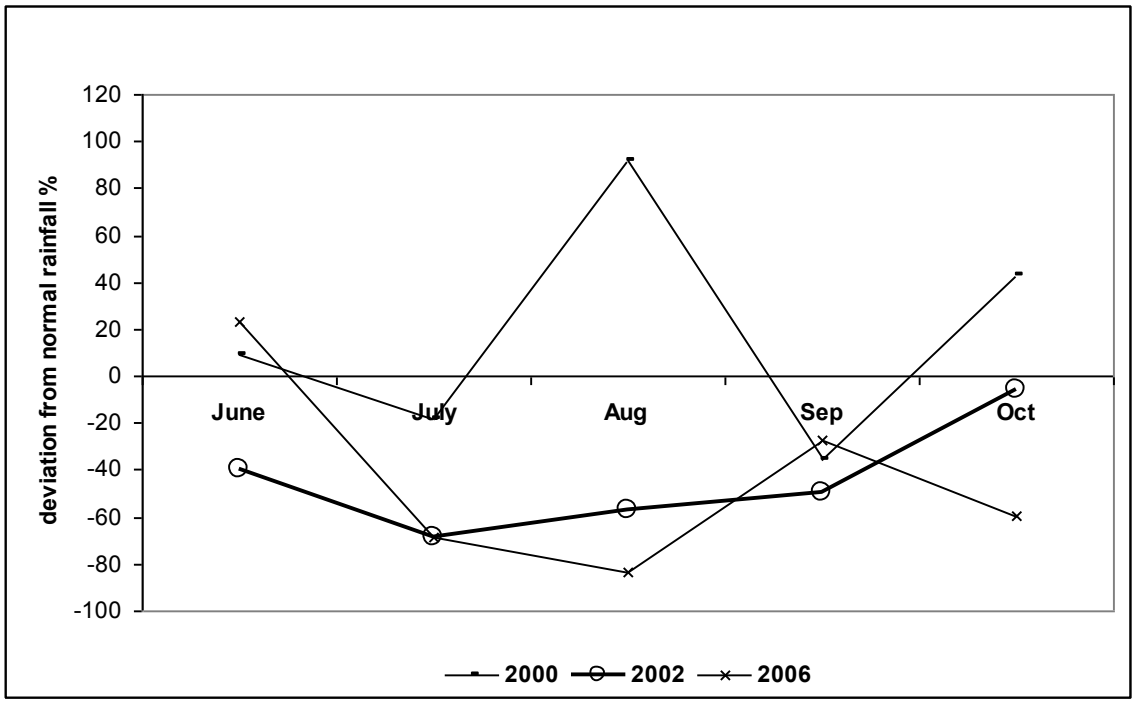

**Figure 4 Rainfall Deviation % from June to October for drought years (2002 and 2006) versus normal year (2000)**

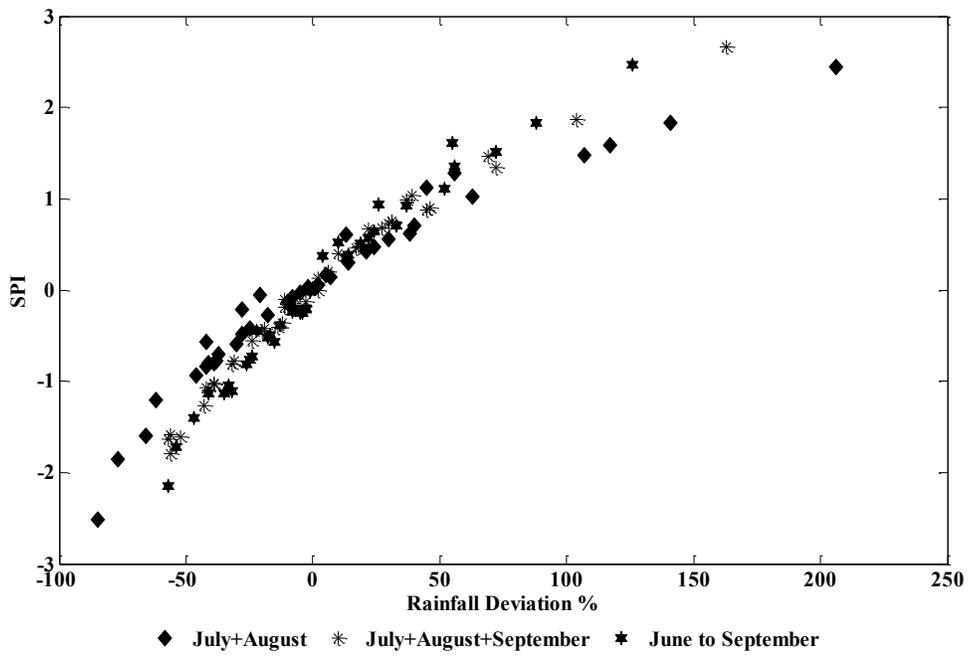

**Figure 5 Scatter plots of SPI versus deviation from normal rainfall for July + August, July + August + September, June to September**

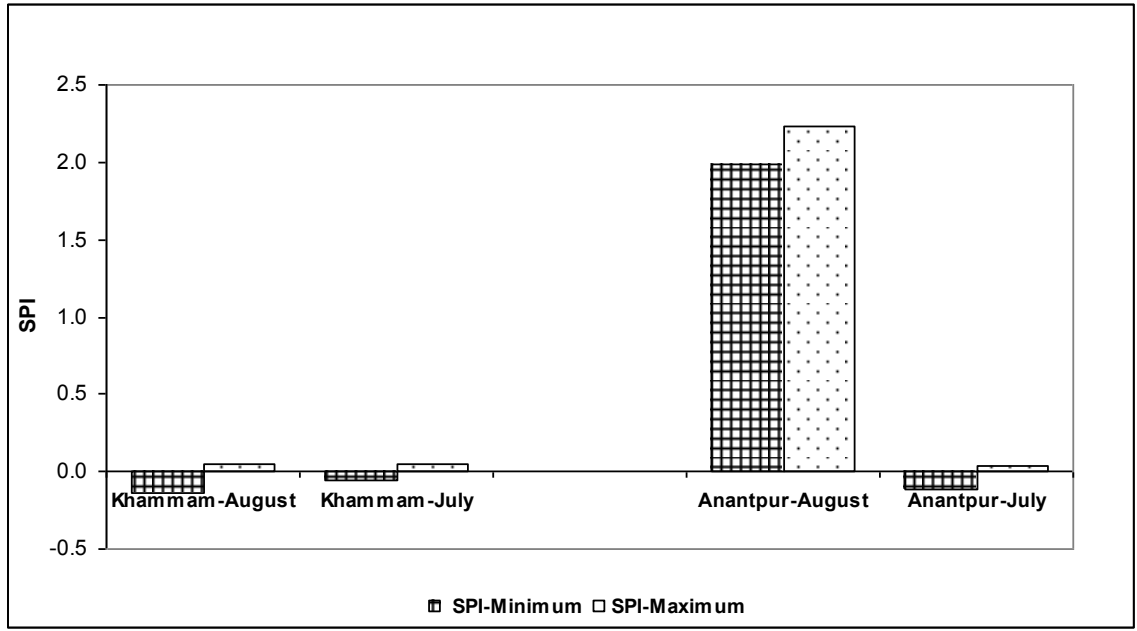

**Figure 6 Range of SPI of different time scales (21 years and 39 years)**

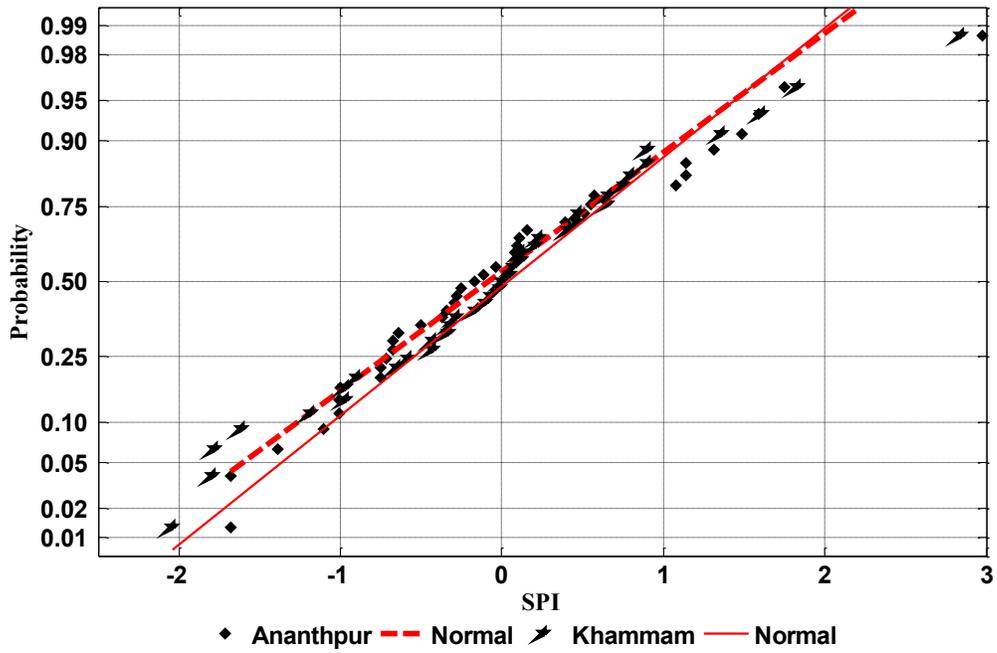

**Figure 7** Normal Probability plot for July SPI

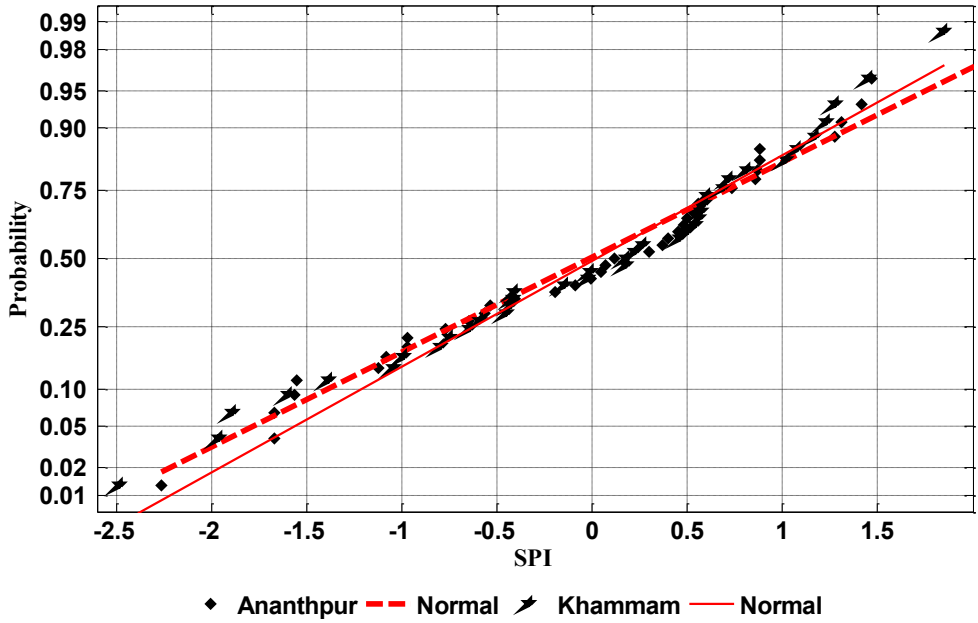

**Figure 8** Normal Probability plot for August SPI

**Table I Drought categories from SPI (Source: Mc Kee et al., 1993)**

| SPI | Drought category |
|---|---|
| 0 to -0.99 | Mild drought |
| -1.00 to -1.49 | Moderate drought |
| -1.5 to -1.99 | Severe drought |
| -2.00 or less | Extreme drought |

**Table II Rainfall pattern in the study area districts**

| Month | District wise normal rainfall (mm) | |
|---|---|---|
| | Ananthpur | Khammam |
| June | 64 | 132 |
| July | 67 | 314 |
| August | 89 | 280 |
| September | 118 | 165 |
| October | 111 | 106 |
| Total | 449 | 997 |

**Table III Very low rainfall events not associated with a very low SPI and very high rainfall not associated with a very high SPI**

| Month | Year | Actual rainfall (mm) | Rainfall deviation from normal % | SPI |
|---|---|---|---|---|
| June | 1988 | 13 | -72 | -1.694 |
| | 1984 | 15 | -68 | -1.532 |
| | 2004 | 18 | -72 | -1.316 |
| | 2001 | 19 | -70 | -1.250 |
| | 1987 | 84 | 79 | 1.057 |
| | 1991 | 131 | 179 | 2.007 |
| | 2007 | 141 | 130 | 2.175 |
| | 1996 | 145 | 209 | 2.247 |
| July | 1997 | 10 | -82 | -1.676 |
| | 1972 | 14 | -74 | -1.389 |
| | 1991 | 19 | -66 | -1.107 |
| | 1984 | 137 | 158 | 1.489 |
| | 2005 | 145 | 116 | 1.592 |
| | 1988 | 158 | 182 | 1.753 |
| | 1989 | 280 | 400 | 2.977 |
| August | 1972 | 6 | -92 | -2.263 |
| | 1984 | 13 | -84 | -1.671 |
| | 2004 | 15 | -83 | -1.562 |
| | 1969 | 156 | 98 | 1.308 |
| | 1998 | 166 | 131 | 1.418 |
| | 2000 | 171 | 92 | 1.471 |
| September | 1969 | 27 | -80 | -2.028 |
| | 1994 | 30 | -75 | -1.908 |

|  | 2003 | 46 | -61 | -1.384 |
|  | 1974 | 231 | 75 | 1.299 |
|  | 2001 | 244 | 107 | 1.418 |
|  | 1988 | 265 | 117 | 1.602 |
|  | 1981 | 283 | 114 | 1.753 |
| October | 1976 | 39 | -58 | -1.495 |
|  | 1997 | 43 | -55 | -1.366 |
|  | 1991 | 197 | 105 | 1.234 |
|  | 1989 | 208 | 124 | 1.353 |
|  | 2001 | 226 | 104 | 1.541 |
|  | 1975 | 248 | 167 | 1.757 |

**Table IV SPI of longer time scales in drought and normal years**

| Year | Situation on ground | June+ July | July+ August | August+ September | June+ July+ August | July+ August+ September | June to September |
|---|---|---|---|---|---|---|---|
| 2000 | Normal | 0.352 | 1.126 | 0.538 | 1.293 | 0.403 | 0.522 |
| 2002 | Drought | -1.173 | -1.2 | -1.424 | -1.407 | -1.635 | -1.717 |
| 2006 | Drought | -0.145 | -1.85 | -1.358 | -1.063 | -1.582 | -1.135 |

**Table V Measured values of parameters for testing normality of SPI from June to October in the study area districts**

| District | Parameters Measured | June | July | August | September | October |
|---|---|---|---|---|---|---|
| Anantpur | w value | 0.954 | 0.956 | 0.973 | 0.973 | 0.961 |
|  | p value | 0.170 | 0.170 | 0.165 | 0.165 | 0.168 |
|  | median | 0.023 | 0.171 | 0.117 | 0.109 | 0.085 |
| Khammam | w value | 0.948 | 0.978 | 0.966 | 0.969 | 0.981 |
|  | p value | 0.172 | 0.164 | 0.167 | 0.166 | 0.163 |
|  | median | 0.041 | 0.003 | 0.182 | 0.188 | 0.057 |